\documentclass[conference]{IEEEtran}
\IEEEoverridecommandlockouts
\usepackage{cite}
\usepackage{amsmath,amssymb,amsfonts}
\usepackage{algorithmic}
\usepackage{graphicx}
\usepackage{textcomp}
\usepackage{amsmath,bm}
\usepackage{xcolor}
\usepackage{geometry}
\usepackage{float}
\usepackage{stfloats}
\usepackage{xcolor}
\usepackage{enumitem}

\usepackage{amsthm}
\theoremstyle{definition}

{
      \theoremstyle{plain}
      
  }
\usepackage{colortbl,booktabs}
\usepackage{caption}
\usepackage{subcaption} 
\def\BibTeX{{\rm B\kern-.05em{\sc i\kern-.025em b}\kern-.08em
    T\kern-.1667em\lower.7ex\hbox{E}\kern-.125emX}}
\begin{document}

\title{Economic Dispatch and Power Flow Analysis for Microgrids}




\author{\IEEEauthorblockN{Saskia Putri\textsuperscript{1}, Xiaoyu Ge$^{2}$, Javad Khazaei\textsuperscript{3}}\\
\IEEEauthorblockA{\textit{$^{(1)}$ Dep. of Civil and Environmental Engineering, $^{(2),(3)}$ Dep. of Electrical and Computer Engineering}\\
\textit{Lehigh University},
Bethlehem PA, USA \\
Emails: \textit{sap322@lehigh.edu, xig620@lehigh.edu, khazaei@lehigh.edu}}
\thanks{This research was in part under support from the Department of Defense (DoD), Office of Naval Research (ONR) award number N00014-23-1-2602 and in part under support from the DoD-ONR award number N00014-23-1-2402.}
}
\maketitle

\begin{abstract}
This study investigates the economic dispatch and optimal power flow (OPF) for microgrids, focusing on two configurations: a single-bus islanded microgrid and a three-bus grid-tied microgrid. The methodologies integrate renewable energy sources (solar PV and wind turbines), battery energy storage systems (BESS), and conventional generators (CHP, diesel, and natural gas), which are connected to the grid to ensure cost-efficient and reliable operation. The economic dispatch analysis evaluates the allocation of generation resources over daily and weekly horizons, highlighting the extensive utilization of renewable energy and the strategic use of BESS to balance system dynamics. The OPF analysis examines the distribution of active and reactive power across buses while ensuring voltage stability and compliance with operational constraints. Results show that the microgrid consistently satisfies load demand with minimal reliance on costly external grid power. Renewable energy sources are maximized for cost reduction, while BESS is employed strategically to address renewable intermittency. For the grid-tied microgrid, optimal power dispatch prioritizes cheaper sources, with Bus 1 contributing the largest share due to its favorable cost profile. Voltage variations remain within acceptable boundaries but indicate potential stability challenges under dynamic load changes, suggesting the need for secondary voltage control. These findings demonstrate the effectiveness of the proposed methodologies in achieving sustainable, cost-effective, and stable microgrid operations.
\end{abstract}

\section{SYSTEM DESCRIPTION}\label{sec:WDS}
\subsection{System Configurations}
\begin{figure}[ht!]
    \centering
    \begin{subfigure}[a]{\columnwidth}
        \centering
        \includegraphics[width=0.6\columnwidth]{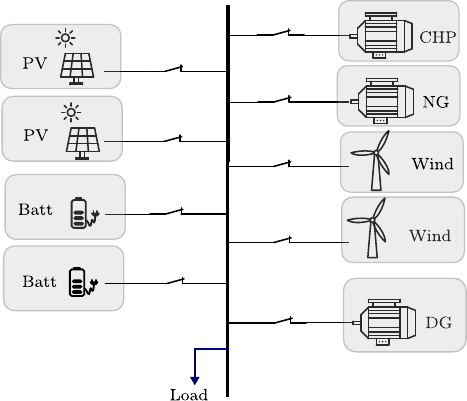}
        \caption{Single-bus system (Islanded Microgrid).}
        \label{fig:sb}
    \end{subfigure}
    \begin{subfigure}[b]{\columnwidth}
        \centering
        \includegraphics[width=0.6\columnwidth]{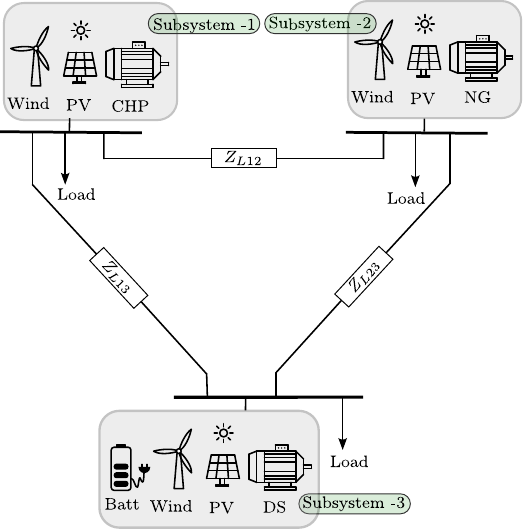}
        \caption{Three-bus system (Grid-Tied Microgrid).}
        \label{fig:3b}
    \end{subfigure}
    \caption{System configurations studied: (a) Single-bus islanded microgrid, and (b) Three-bus grid-tied microgrid.}
    \label{fig:system_configurations}
\end{figure}

In this study, we analyze the operation of two microgrids (MG) configurations: a single-bus system representing a single-bus MG and a three-bus system representing a grid-tied MG. The single-bus system is designed to operate independently of the main grid, relying entirely on its internal generation units, including renewable energy sources (RESs), battery energy storage systems (BESS), and conventional generators (DG), to meet load demand and maintain system stability. The structure of the single-bus MG is illustrated in Fig.~\ref{fig:sb}.

On the other hand, the three-bus system introduces a more complex topology, integrating the MG with the main power grid, displayed in Fig.~\ref{fig:3b}. This configuration allows for power exchange between the MG and the grid, leveraging dynamic pricing and external support to optimize costs while maintaining reliability.
\subsection{Load Profile}
\begin{figure}[t!]
 \vspace{-0.5cm}
    \centering
    \includegraphics[width = 1\columnwidth]{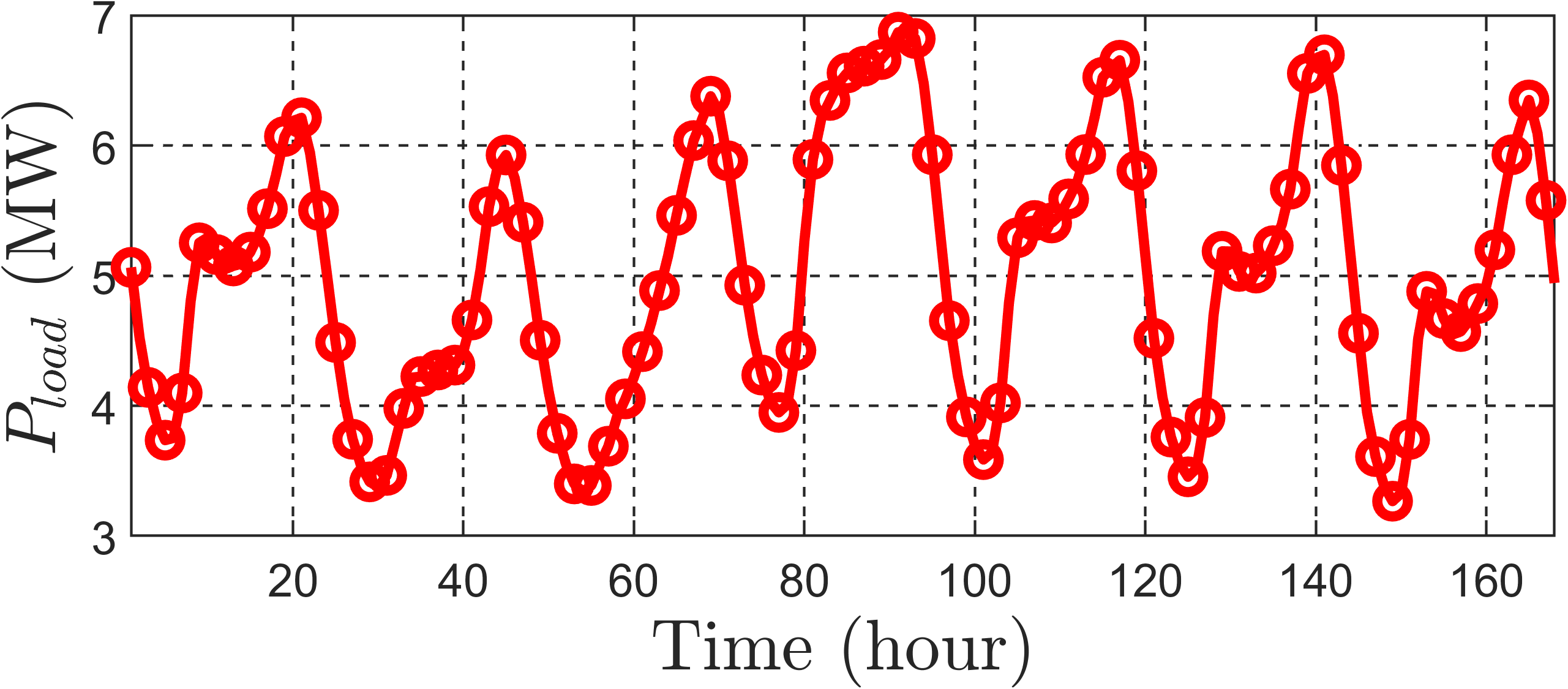}
    \caption{Normalized load profile for New York State}
    \label{fig:load}
     \vspace{-0.1cm}
\end{figure}

Load profiles represent the system's energy demand over time. In this study, the grid-tied microgrid accommodates partial load requirements from New York State ($P_{load}(t)$) with a mean value of 5MW. The aggregated load profile is obtained from~\cite{EIA_NewYork_Data}. A normalized load profile of around 5MW using the following normalization function:
\begin{equation}
    P^n_{load}(t)=\frac{P^o_{load}(t)-\bar{P^o_{load}(t)}}{\sigma(P^o_{load}(t))}+5
\end{equation}
where $P^n_{load}(t)$ and  $P^o_{load}(t)$ are the normalized and original hourly load, while $\bar{P^o_{load}(t)}$ and $\sigma(P^o_{load}(t))$ are the mean and the standard deviation of the original load. The normalized hourly load profile is displayed in Fig.~\ref{fig:load}.
\subsection{Elecricity Price}
\begin{figure}[t!]
 \vspace{-0.3cm}
    \centering
    \includegraphics[width = 1\columnwidth]{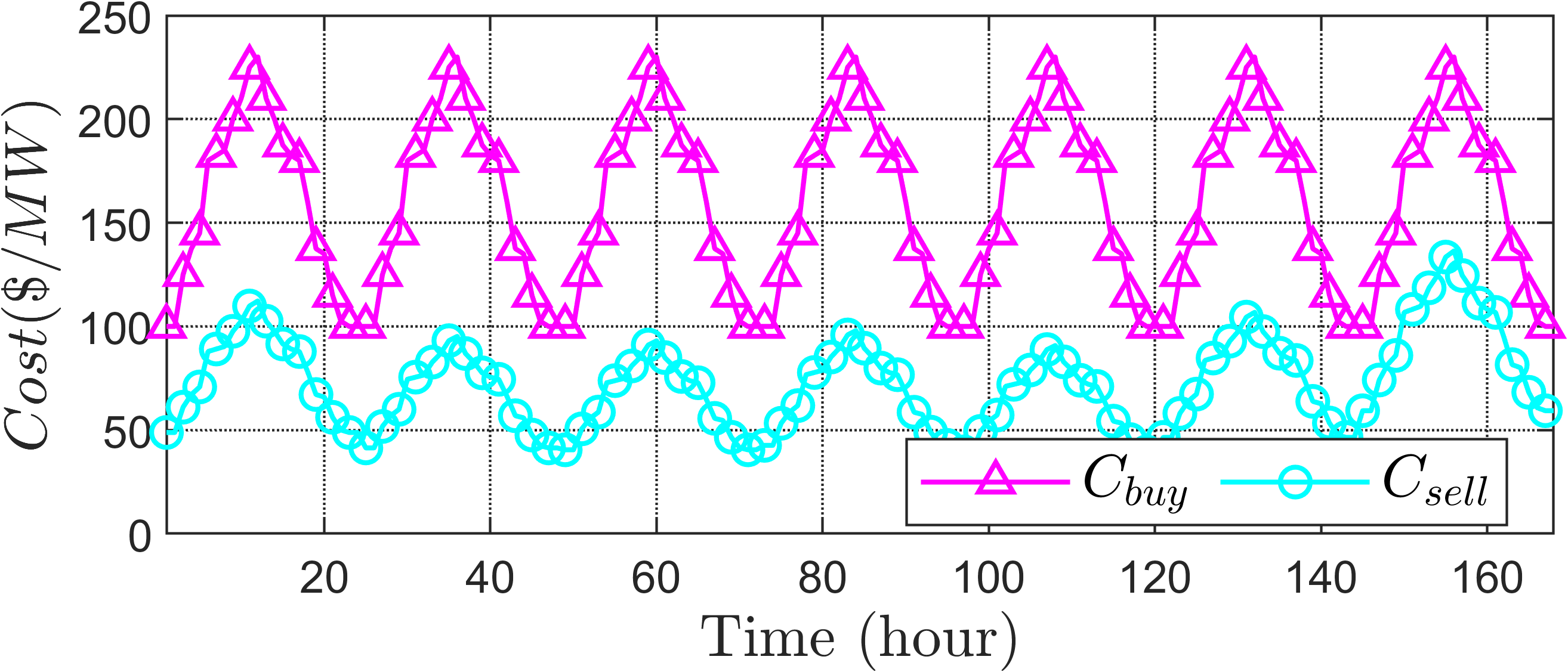}
    \vspace{-0.5cm}
    \caption{Wholesale electricity price}
    \label{fig:price}
\end{figure}

Given that the microgrid is designed to be grid-connected, the objective function of the optimization also includes minimization of the grid power dependent on the dynamic price. Here, we use the wholesale electricity price from \cite{EIAWholesaleElectricity} from September 15 - 21, 2023 with a pattern from \cite{wang2020receding} and with a 1:0.8 ratio between the buying and selling price. The hourly dynamic pricing is exhibited in Fig.~\ref{fig:price}.  
\subsection{Conventional Generators (CG)}
 Conventional generators serve as dependable power sources, capable of providing consistent energy output irrespective of external conditions. In this study, three conventional generation units are considered: 1) a combined heat and power (CHP) plant, 2) a diesel generator (DS), and 3) a natural gas (NG) generator. Their economic dispatch is determined based on cost coefficients~\eqref{costdg} and ramping capabilities~\eqref{uc2}-\eqref{uc3} to ensure cost-effectiveness while meeting load requirements. 

 The operation cost of each distributed generation (DG) unit relates to the dispatched power, maintenance costs, and fuel cost coefficient, mathematically expressed as~\cite{moazeni2020dynamic}:
  \begin{align}
  \begin{cases}
       C_{CHP}(t)&=a_{CHP}+b_{CHP} \cdot P_{CHP}(t)+c_{CHP} \cdot P_{CHP}(t)^2 \\
      C_{NG}(t)&=a_{NG}+b_{NG} \cdot P_{NG}(t)+c_{NG} \cdot P_{NG}(t)^2 \\
      C_{DS}(t)&=a_{DS}+b_{DS} \cdot P_{DS}(t)+c_{DS} \cdot P_{DS}(t)^2 
  \end{cases} \label{costdg}
\end{align}
where $a_{CGi}, b_{CGi}, c_{CGi}$ are the cost coefficients of the $i$-th CG. 
 
 To represent the real-world operation of the generators, several constraints are imposed, as follows~\cite{moazeni2021step}:
 \begin{align}
  &  U_{CG,i}(t)\underline{P}_{G,i}\leq P_{G,i}(t)\leq U_{CG,i}(t)\overline{P}_{G,i}  \label{uc1}\\
  &  P_{G,i}(t)-P_{G,i}(t-1)\leq \overline{R}_{up}U_{CG,i}(t-1)  \label{uc2}\\
 &   P_{G,i}(t-1)-P_{G,i}(t)\leq \overline{R}_{down}U_{CG,i}(t)  \label{uc3}\\
 & S_{up,i}(t)=\mathrm{Max}(0,U_{CG,i}(t)-U_{CG,i}(t-1)), \label{uc4} \\
  & S_{dn,i}(t)=\mathrm{Max}(0,U_{CG,i}(t-1)-U_{CG,i}(t)), \label{uc5} \\
  &\forall i\in CG, \forall t\in T  \nonumber
\end{align}
where \eqref{uc1} limit the output power for each CG, $U_{CG,i}(t)\ \in\{0,1\}$ represent on/off status of the $i$-th CG unit $i$. Equations~\eqref{uc2} and~\eqref{uc3} denote the ramp-up and ramp-down limits of the $i$-th CG, respectively while $S_{up,i}(t)$ and $S_{dn,i}(t) \in\{0,1\}$ in~\eqref{uc4} and~\eqref{uc5} denote the start-up and shut down status of $i$-th CG unit at time $t$, respectively. 
\subsection{Battery Energy Storage System(BESS)}
Battery energy storage systems play a critical role in balancing the intermittency of renewables and ensuring the reliability of the microgrid. In this study, two BESSs are used to balance the energy. Their operation is optimized to minimize costs while maintaining state-of-charge constraints and operational limits, expressed as follows~\cite{bordons2020model}:
\begin{align}
    E_{Bi}(t) &=  E_{Bi}(t-1)+  \eta P_{Bi}(t) \Delta t \label{ebat} \\
    -\overline{P}_{Bichg} &\leq P_{Bi}(t)\leq \overline{P}_{Bidis} \label{boundbat} \\
     \underline{E}_{Bi} &\leq E_{Bi}(t)\leq \overline{E}_{Bi} \label{boundeb} \forall i\in \mathcal{N}_B, \forall t\in T 
\end{align}
where \eqref{ebat} denotes the dynamic changes between the available energy of the $i$-th BESS ($E_{Bi}(t)$) and its power output ($P_{Bi}(t)$) multiplied by the efficiency $\eta$. Equations \eqref{boundbat}-\eqref{boundeb} limit the output power and the energy of the BESSs, respectively.
\subsection{Renewable Energy Sources}
Renewable energy sources are pivotal to achieving sustainability within the microgrid. In this study, two renewable sources are utilized, consisting of two wind turbines and two solar PV generation units. These sources are prioritized for their cost-effectiveness and environmental benefits, yet their intermittency necessitates complementary generation strategies. Their power output heavily relies on weather data such as temperature, solar radiation, and wind speed. In this study, weather data is obtained from~\cite{VisualCrossingWeather} on September 15 - 21, 2023. 
\subsubsection{Wind Turbine Power Generation}
\begin{figure}[t!]
 \vspace{-0.3cm}
    \centering
    \includegraphics[width = 1\columnwidth]{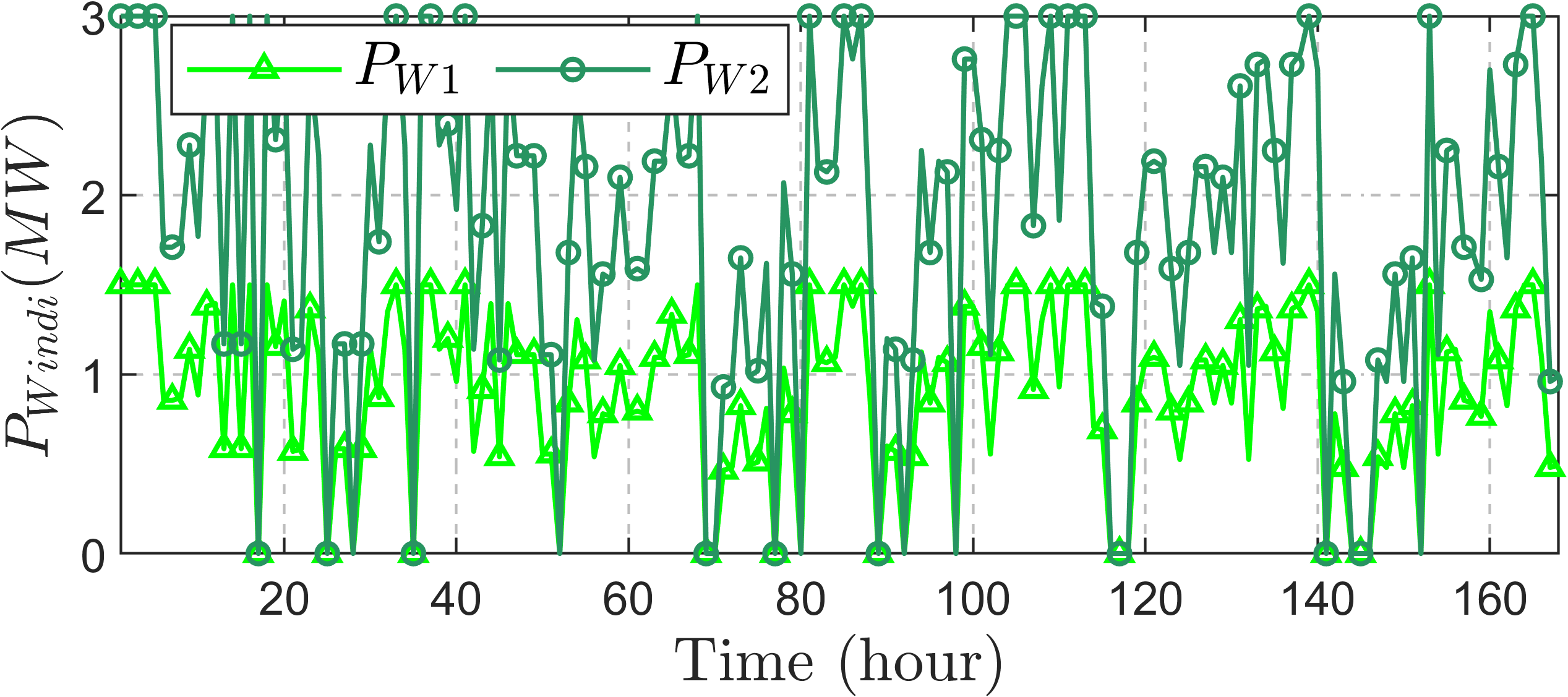}
    \caption{Hourly power output of Wind turbine}
    \label{fig:WT}
     \vspace{-0.5cm}
\end{figure}

The  power output is defined as~\cite{moazeni2021step}:
\begin{align}
    P_{WT} (t) &=
\begin{cases}
0 & \mathrm{if} \upsilon_t < \upsilon_{i} \\
P_{r}\frac{\upsilon_t-\upsilon_{i}}{\upsilon_{r}-\upsilon_{i}} & \mathrm{if} \ \upsilon_{i}\leq \upsilon_t\leq \upsilon_{r}\\
P_{r} & \mathrm{if} \ \upsilon_{r}\leq \upsilon_t\leq \upsilon_{o} \\
0 & \mathrm{if} \upsilon_t > \upsilon_{o} \label{P_wind}
\end{cases}
\end{align}
where $\upsilon_{t}$ and $\upsilon_{r}$ are the wind speed at time instant $t$ and the rated wind speed, respectively in $m/s$, $P_{r}$ is the rated power in $kW$, $\upsilon_i$ and $\upsilon_o$ are the cut-in and cut-out wind speed in $m/s$. The hourly power output of the Wind Turbine over a week is illustrated in Fig.~\ref{fig:WT}.
\subsubsection{Solar Photovoltaic (PV) Power Generation}
\begin{figure}[t!]
 \vspace{-0.5cm}
    \centering
    \includegraphics[width = 1\columnwidth]{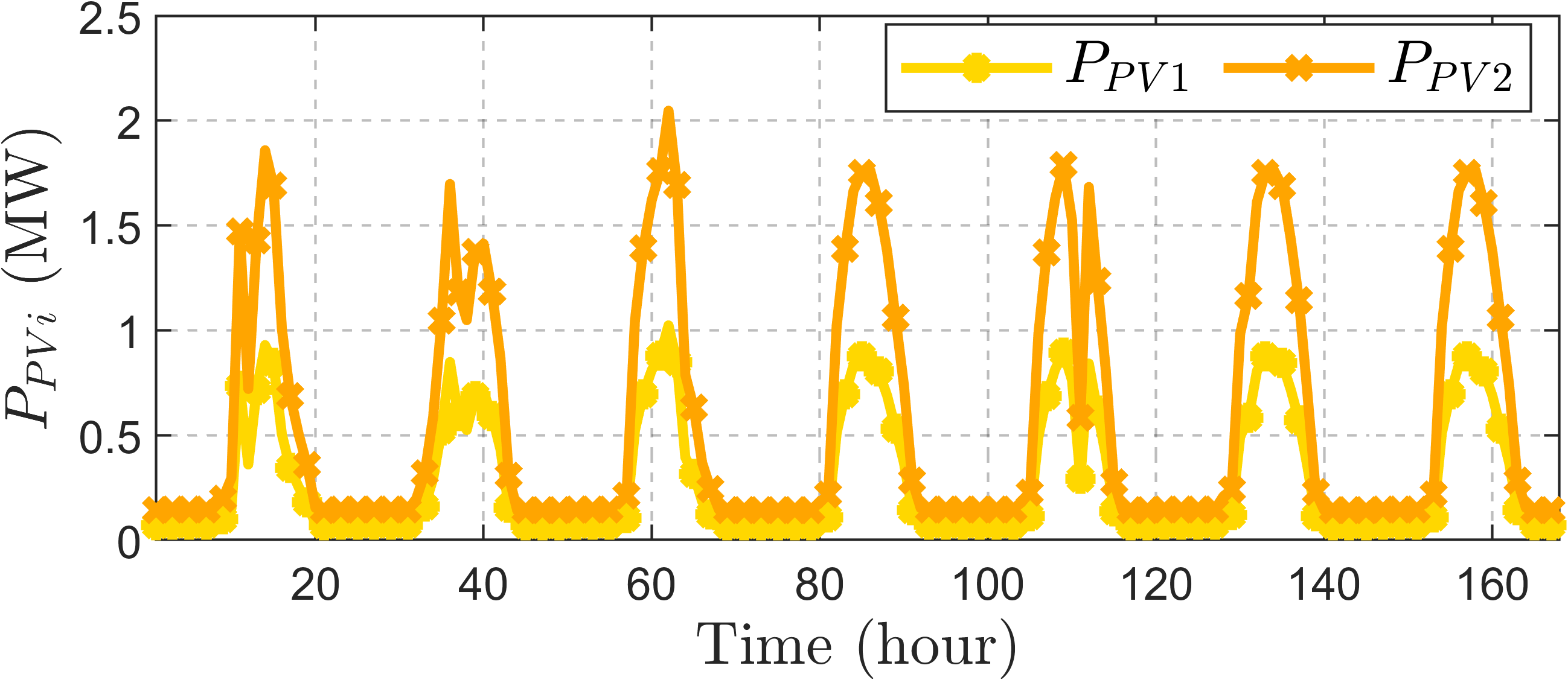}
    \caption{Hourly power output of solar PV}
    \label{fig:PV}
     \vspace{-0.5cm}
\end{figure}

The PV power output can be modeled as~\cite{hijjo_pv-battery-diesel_2017}:
\begin{equation}
   P_{i,PV}(t)=P_{STC}\dfrac{G_{ac}(t)}{G_{STC} }(1+K(T_c-T_i)) \label{p_pv}
\end{equation}
where $P_{STC}$ is the power output of PV in standard temperature condition (STC), $G_{ac}(t)$ expressed as the solar irradiance at time instant $k$, and $K$ is the temperature coefficient of the PV unit ($-0.0047$) \cite{hijjo_pv-battery-diesel_2017}. $T_c$ and $T_i$ are the reference (set at $25^{\circ}$C) and real temperature of solar cells in $^{\circ}$C. The hourly power output generation profile of the solar PV is depicted in the top left subplot in Fig.~\ref{fig:PV}.
\subsection{Power Balance}
To ensure optimal allocation of energy resources, the total generated power must satisfy the load requirement from~\eqref{balance}. 
\begin{align}
P_{load}(t) &= P_{Grid}(t) + \sum_{t\in N_{CG}}P_{CGi}(t) \nonumber \\
&+ \sum_{i \in \mathcal{N}_{PV}} P_{PVi}(t)+\sum_{i \in \mathcal{N}_{WT}}P_{WTi}(t) + \sum_{i \in \mathcal{N}_{B}}P_{B}(t)  \label{balance}
\end{align}
where the summation of generated powers by the grid, DGs, PV, wind, and BESS should be equal to the consumed power. 
\subsection{Optimal Power Flow (OPF)}
The optimal power flow problem is one of the most common and important problems that need to be solved for microgrid operations. The optimal power flow problem aims to solve the optimal power distribution problem to minimize the total cost on the transmission level based on the results from economic dispatch. The detailed configuration of the grid-tied MG displayed in Fig.\ref{fig:3b} is as follows:
\begin{enumerate}
    \item A three PQ bus system including three conventional generators, two PVs, two wind turbines, and two BESSs model is considered with the following configurations:
    \begin{itemize}
        \item $Bus \ 1 =  P_{CHP} + P_{PV1} + P_{WT1}$
        \item $Bus \ 2 =  P_{NG} + P_{PV2} + P_{B1}$
        \item $Bus \ 3 =  P_{DS} + P_{PWT2} + P_{B2}$
    \end{itemize}
    \item Line parameters to be purely inductive with impedance $Z = j1 \Omega$.
    \item We assume the reactive power load ($Q_i(t)$) for the $i$-th bus is 20\% of the active power load.
\end{enumerate} 
The OPF problem is subject to a variety of constraints, which ensure the feasible operation and stability of the microgrid. These constraints include:
\subsubsection{Bus power balance constraints}\mbox{} \\
The active and reactive power injected at each bus must satisfy the load demand while accounting for network losses~\cite{power}:
\begin{align} 
P_{l}&=\sum_{i=1}^{n}P_{G,i,l}-\sum_{k=1}^{m}P_{D,l,k}\nonumber\\
&=|V_{l}|\sum_{j=1}^{N}|V_{j}||Y_{i,j}|cos(\theta_{i,j}-\delta_{i}+\delta_{j})\label{Pi} \\
Q_{l}&=\sum_{i=1}^{n}Q_{G,i,l}-\sum_{k=1}^{m}Q_{D,l,k}\nonumber\\
&=-|V_{l}|\sum_{j=1}^{N}|V_{j}||Y_{l,j}|sin(\theta_{l,j}-\delta_{l}+\delta_{j}),  \label{Qi}\\ 
&\forall l\in \mathcal{N}_{Bus}, \forall t\in T   \nonumber
\end{align}
These equations describe the balance of active and reactive power at each bus in the system with \(P_{l}\) representing the total active power injected at bus \(l\), calculated as the difference between the total active power generated, \(\sum_{i=1}^{n}P_{G,i,l}\), and the total active power demand, \(\sum_{k=1}^{m}P_{D,l,k}\). Similarly, \(Q_{l}\) denotes the total reactive power injected at bus $l$, which is the difference between the total reactive power generated, \(\sum_{i=1}^{n}Q_{G,i,l}\), and the reactive power demand, \(\sum_{k=1}^{m}Q_{D,l,k}\). 

The terms \(|V_{l}|\) and \(|V_{j}|\) represent the voltage magnitudes at buses \(l\) and \(j\), respectively, while \(|Y_{i,j}|\) denotes the magnitude of the admittance of the transmission line connecting buses \(i\) and \(j\). The phase angle of the line admittance is \(\theta_{i,j}\), and \(\delta_{i}\) and \(\delta_{j}\) are the voltage phase angles at buses \(i\) and \(j\), respectively. The cosine and sine terms, \(\cos(\theta_{i,j} - \delta_{i} + \delta_{j})\) and \(\sin(\theta_{i,j} - \delta_{i} + \delta_{j})\), represent the real and imaginary components of the power flow between buses \(i\) and \(j\), respectively. 

The variables \(P_{l}\) and \(Q_{l}\) are evaluated for all buses \(l \in \mathcal{N}_{Bus}\) and for all time steps \(t \in T\). These equations ensure that the active and reactive power injected at each bus is balanced with the corresponding generation, load demand, and network effects.

\subsubsection{Bus Power Limits}\mbox{} \\
Each generator in the $i$-th bus operates within its minimum and maximum output capacity~\cite{power}: 
\begin{align} \underline{P}_{i,l} &\leq P_{i,l}(t) \leq \overline{P}_{i,l} \label{pbusbound} \\
 \underline{Q}_{i,l} &\leq Q_{i,l}(t) \leq \overline{Q}_{i,l} \label{qbusbound}\\ 
 &\forall i \in \mathcal{N}_{G}, \ \forall l \in \mathcal{N}_{Bus}, \ \forall t\in T. \nonumber
 \end{align}
 where $\mathcal{N}_G$ defines the generator set in this study. 
 \subsubsection{Bus Voltage Constraints}\mbox{} \\
The voltage magnitude at each bus must remain within specified bounds to ensure system stability:
 \begin{align} \underline{V}_{i} &\leq V_{i}(t) \leq \overline{V}_{i}, \quad \forall i \in \mathcal{N}_{Bus}.  \label{vbusbound} 
 \end{align}
 \subsubsection{Voltage Phase Angle Limits}\mbox{} \\
 To ensure that the voltage phase angle $\delta_{i}(t)$ remains within physically and operationally realistic limits, the following limits are applied~\cite{power}:
 \begin{align} -\pi &\leq \delta_{i}(t)  \leq \pi, \ \ \forall i\in \mathcal{N}_{Bus}, \forall t\in T. \label{piband}
 \end{align}
\subsubsection{System Parameters}
All parameters used to perform the economic dispatch of the grid-connected microgrid can be seen in Table.~\ref{tab:param}. Cost coefficients are different for each generator which will be reflected in their contribution. Additionally, all initial conditions are set to be 0 except the energy of the batteries at its maximum capacity. 
\begin{table*}[t!]
\caption{Parameters of the grid-connected microgrid}
\label{tab:param}
\resizebox{\textwidth}{!}{%
\begin{tabular}{@{}cccccccccll@{}}
\toprule
Parameters & Qty & Unit & Parameters & Qty & Unit & Parameters & Qty & Unit &  &  \\ \midrule
\multicolumn{3}{c}{Combined Heat and   Power} & \multicolumn{3}{c}{BESS 1} & \multicolumn{3}{c}{Wind turbines} &  &  \\ \cmidrule(r){1-9}
$a,b,c$ & 1530,0.010,0.000233 &  & $C_B$ & 0.025 \$/kWh&  & $P_{WT,1},P_{WT,2}$ & 1.5, 3 & MW &  &  \\
$\overline{P}_{CHP}$ & 6 & MW & $P_{B1}$ & 4 & MW & $v_i$ & 5 & $m/s$ &  &  \\
$\overline{R}_{up}$, $\overline{R}_{dn}$ & 0.6 & MW & $E_{B1}$ & 8 & MWh & $v_{nom}$ & 15 & $m/s$ &  &  \\ \cmidrule(r){1-3}
\multicolumn{3}{c}{Diesel Generators} & $\underline{E}_{B1}$ & 0.15 $E_{B1}$ &  & $v_o$ & 45 & $m/s$ &  &  \\ \cmidrule(r){1-6}
$a,b,c$ & 1300, 0.013, 0.000235 &  & \multicolumn{3}{c}{BESS 2} & $C_{WT}$ & 0.0018 \$/kWh&  &  &  \\ \cmidrule(lr){4-9}
$\overline{P}_{DS}$ & 4 & MW & $P_{B2}$ & 6 & MW & \multicolumn{3}{c}{Solar PVs} & \multicolumn{1}{c}{} & \multicolumn{1}{c}{} \\ \cmidrule(lr){7-9}
$\overline{R}_{up}$, $\overline{R}_{dn}$  & 0.5 & MW & ${E}_{B2}$ & 12 & MWh & $P_{PV,1},P_{PV,2}$ & 1,2 & MW &  &  \\ \cmidrule(r){1-3}
\multicolumn{3}{c}{Natural Gas} & $\underline{E}_{B2}$ & 0.15 $E_{B2}$ & MWH & $C_{PV}$ & 0.0025 \$/kWh&  &  &  \\ \cmidrule(r){1-3} \cmidrule(lr){7-9}
$a,b,c$ & 992, 0.016. 0.000241 &  & $\eta$ & 95 & \% & \multicolumn{3}{c}{Simulation setup} &  &  \\ \cmidrule(lr){4-9}
$\overline{P}_{NG}$ & 10 & MW & \multicolumn{3}{c}{$P_{Gr}$} & $T$ & 168 & hr &  &  \\ \cmidrule(lr){4-6}
$\overline{R}_{up}$, $\overline{R}_{dn}$ & 1 & MW & $\bar{C}_{buy}$ & 31.91 & \$/MWh & $\Delta t$ & 1 & hr &  &  \\
Cup \&dn & 0.001 & $\$/MWh$ & $\bar{C}_{sell}$ & 30.18 & \$/MWh & nvar & 21 & vars &  &  \\ \bottomrule
\end{tabular}%
}
\end{table*}
\section{PROBLEM FORMULATION}\label{sec:mpc}
In this study, mixed integer nonlinear programming (MINLP) is utilized to implement the economic dispatch and optimal power flow operation of the islanded and grid-tied microgrid with optimization toolbox from~\cite{Currie_OPTI} ran in MATLAB.
\subsection{Decision Variables} 
In this study, the decision variables of the optimization include all active power from the available sources, including the activation status of each DGU listed as follows
\begin{align}
    X &= \begin{bmatrix}
        P_{CHPi} \ P_{NGi} \ P_{DSi}  \ P_{Bi}  \ P_{PVi}  \ P_{WTi} \ P_{Gr} \nonumber \\ E_{Bi}  \ U_{CGi}  \ S_{dni}  \ S_{upi}
    \end{bmatrix}^{\top}
\end{align}
In the case of optimal power flow optimization, the decision variables are extended to include the reactive power, bus voltage, and phase angle, as follows 
\begin{align}
    X = \begin{bmatrix}
        P_{CHPi} \ P_{NGi} \ P_{DSi}  \ P_{Bi}  \ P_{PVi}  \ P_{WTi} \ P_{Gr} \nonumber \\ E_{Bi}  \ U_{CGi}  \ S_{dni}  \ S_{upi} \ V_i \ P_i \ Q_i \ \delta_i
    \end{bmatrix}^{\top}
\end{align}
where $V_i,  P_i, Q_i, \delta_i$ are the $i$-th bus voltage, active and reactive powers of the bus, and the phase angle. 
\subsection{Objective Function} 
To ensure cost-efficient and demand-driven operation, the composite objective function is divided into the following components, as follows~\cite{moazeni2021step}:
\begin{itemize}
    \item {Grid Power Cost:}
    \begin{align}
    \ell_{Gr}(t) &= \left( \frac{C_{buy}(t) + C_{sell}(t)}{2}P_{Gr}(t) \right. \\ 
    &\left. + \frac{C_{buy}(t) - C_{sell}(t)}{2}|P_{Gr}(t)|\right)
    \end{align}
    \textit{Description:} This term accounts for the cost of purchasing electricity from the grid at time $t$, represented by \(C_{buy}(t)\), and the revenue from selling electricity back to the grid, represented by \(C_{sell}(t)\). The net cost depends on the power exchanged with the grid, \(P_{Gr}(t)\).
    \item {Conventional Generator Costs:}
    \begin{align}
    \ell_{CG,i}(t) &= \left( f_{CG,i}(P_{CG,i}(t)) \right. \\
    &\left. + C_{up,i}S_{up,i}(t) + C_{dn,i}S_{dn,i}(t)) \right) \ \forall i \in \mathcal{N}_{CG}
    \end{align}
    \textit{Description:} This term captures the costs associated with conventional generators:
    \begin{itemize}
        \item $f_{CG,i}(P_{CG,i}(t))$: The fuel cost of operating the $i$-th conventional generator as a function of its output power \(P_{CG,i}(t)\) detailed in \eqref{costdg}.
        \item \(C_{up,i}\) and \(C_{dn,i}\): The startup and shutdown costs of the $i$-th generator.
        \item \(S_{up,i}(t)\) and \(S_{dn,i}(t)\): Binary variables indicating the startup and shutdown status of the $i$-th generator at time $t$.
    \end{itemize}
    \item {Battery Energy Storage System (BESS) Costs:}
    \begin{align}
    \ell_{B,i}(t) =C_{Bi} P_{Bi}(t) \ \forall i \in \mathcal{N}_{B}
    \end{align}
    \textit{Description:} This term represents the cost of operating the BESS at time $t$.
    \item {Photovoltaic (PV) Costs:}
    \begin{align}
    \ell_{PV,i}(t) =  C_{PV,i}(P_{PV,i}(t)) \ \forall i \in \mathcal{N}_{PV}
    \end{align}
    \textit{Description:} This term accounts for any operational costs associated with the output power of solar PV.
    \item {Wind Turbine (WT) Costs:}
    \begin{align}
   \ell_{WT,i}(t) = C_{WT,i} \cdot P_{WT,i}(t) \ \forall i \in \mathcal{N}_{WT}
    \end{align}
    \textit{Description:} This term represents costs associated with the output power of wind turbines.
\end{itemize}

\subsection{OPTIMIZATION FORMULATIONS}
Combining all constraints imposed for each generation unit, the OPF, and the objective function for each generator, the economic dispatch problem formulation is as follows
\begin{subequations} \label{eq:wesocp}
            \begin{align}
            \min_{X} \quad &  \sum_{t \in T}\left( \ell_{Gr}(t) + \sum_{i \in \mathcal{N}_{CG}} \ell_{CG,i}(t) \right. \\
            &\left. +   \sum_{i \in \mathcal{N}_{B}} \ell_{B,i}(t) + \sum_{i \in \mathcal{N}_{PV }} \ell_{PV,i}(t) + \sum_{i \in \mathcal{N}_{WT}} \ell_{WT,i}(t) \right) \\ 
                \text{s.t.} \quad &  \eqref{uc1} \ \text{ Generator power limits} \label{c1}\\
                &\eqref{uc2}-\eqref{uc5} \ \text{ Ramping constraints} \\
                &\eqref{ebat}-\eqref{boundeb} \ \text{ Energy storage constraints} \label{c2}\\
                &\eqref{balance} \ \text{Power balance constraint} \label{c3}\\
                &\eqref{Pi}-\eqref{Qi} \ \text{Bus power balance} \label{c4}\\
                &\eqref{pbusbound}-\eqref{qbusbound} \ \text{Active and reactive power limits} \label{c5}\\
                &\eqref{vbusbound}-\eqref{piband} \ \text{Bus voltage and phase angle limits} \label{c6}
            \end{align}
 \end{subequations} 
 Note that for islanded MG operation, the constraints are from \eqref{c1}-\eqref{c3} while for the grid-tied MG operation the constraints include \eqref{c1}-\eqref{c3} and \eqref{c4}-\eqref{c6}.  
\section{CASE STUDIES}\label{sec:casestudies}

In this section, two case studies are presented to analyze the performance of the proposed methods under different operational scenarios. The first case study focuses on economic dispatch analysis applied to a single-bus islanded microgrid, which evaluates the cost-efficient allocation of power generation resources to meet the load demand over a specified time horizon. The second case study investigates the OPF analysis applied to a three-bus grid-tied microgrid, which considers the distribution of active and reactive power across the microgrid while satisfying technical and operational constraints. 
\subsection{SINGLE-BUS MICROGRID OPERATION}
In this case study, the economic dispatch optimization is applied to a single MG. The economic dispatch analysis aims to allocate power generation from different units, including renewable energy sources, conventional generators, and battery energy storage systems, to minimize operational costs. This study evaluates the performance of the system over a week time horizon.
\begin{figure*}[t!]
    \centering
    \includegraphics[width = 1\textwidth]{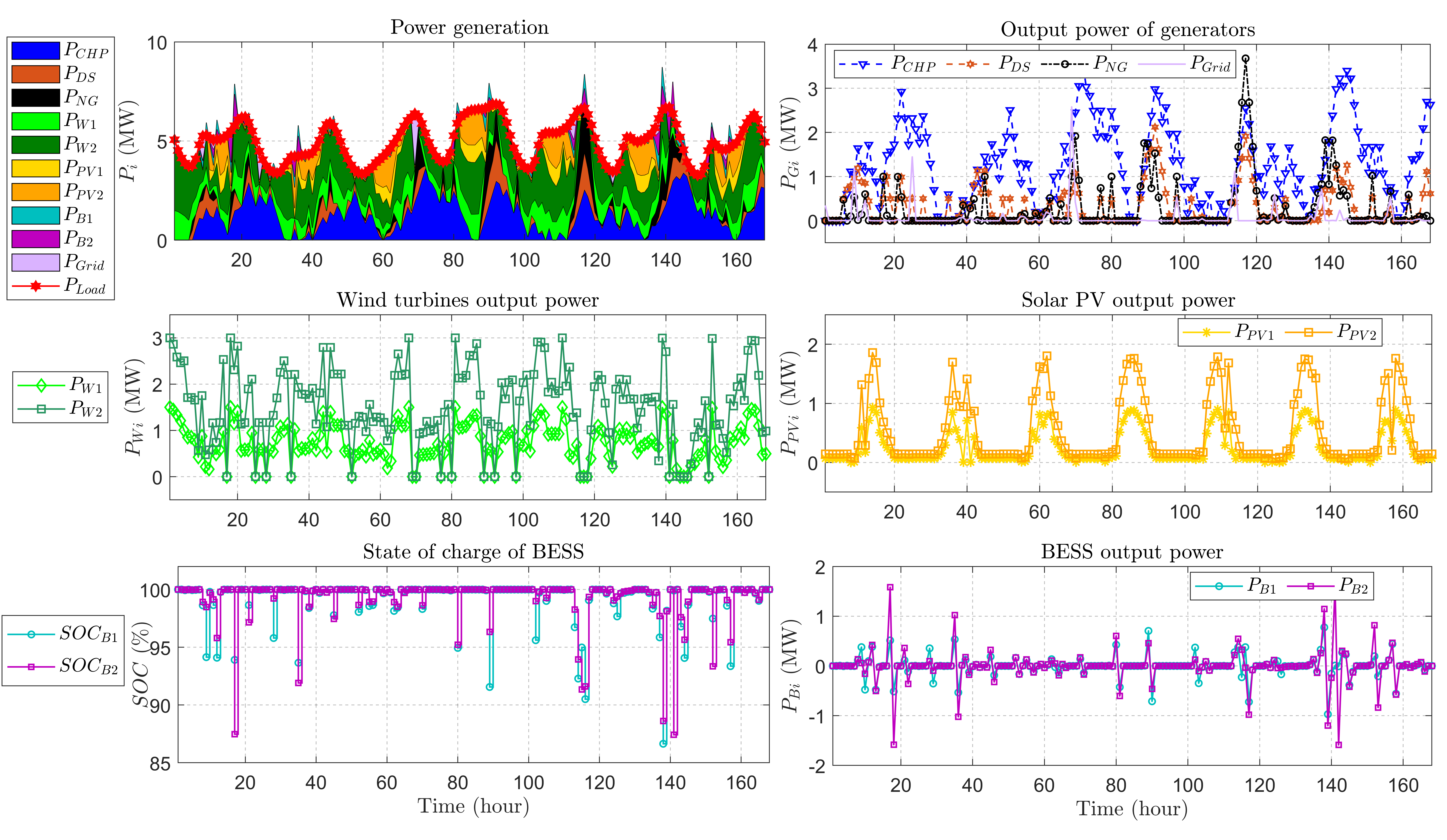}
    \vspace{-0.5cm}
    \caption{Results of the weekly economic dispatch of single-bus MG}
    \label{fig:resultsanr1}
\end{figure*}
The results of the economic dispatch analysis are illustrated in Fig.~\ref{fig:resultsanr1}, showcasing the system's ability to meet load demands while optimizing operational costs. The top left subplot highlights the contributions of each generation unit, where a consistent match between the load line and the cumulative area of power generation confirms that demand is consistently satisfied throughout the dispatch period. This balance demonstrates the effectiveness of optimizing resources to maintain reliability.

Excess power generation is also observed during certain iterations, primarily utilized to charge the Battery Energy Storage System (BESS). The BESS plays a critical role in stabilizing the system by storing surplus energy and restoring its charge to the maximum capacity, as depicted in the bottom subplot of Fig.~\ref{fig:resultsanr1}. This behavior underscores the importance of energy storage in maintaining operational flexibility and enhancing the microgrid's resilience to fluctuating renewable outputs.

The contributions of conventional generators are detailed in the top right subplot. Each generator is utilized intermittently to meet demand, with distinct shutdown periods reflecting their respective roles in the optimization. Among these, the Combined Heat and Power (CHP) unit exhibits the shortest shutdown time (11 instances), followed by the diesel generator (29 instances) and the natural gas generator (38 instances). This staggered operation aligns with the cost and efficiency characteristics of each generator, ensuring that they are dispatched economically.

In contrast, the grid power contribution ($P_{Gr}$), represented by the light purple line, demonstrates a frequent shutdown pattern (75 instances). This is primarily due to the higher costs associated with purchasing power from the grid compared to utilizing the microgrid's internal generation units. This finding highlights the cost-effectiveness of prioritizing local generation and storage over external sourcing.

The optimization also maximizes the use of renewable energy sources, such as wind and solar PV, which is in line with the cost-effectiveness objective. This is evident from the parallel trajectories observed between the input data for wind and solar PV generation (Fig.\ref{fig:PV}-Fig.~\ref{fig:WT}) and their actual power outputs derived from the optimization, as shown in the middle subplot of Fig.\ref{fig:resultsanr1}. This extensive utilization of renewables further emphasizes the microgrid's capability to operate sustainably while minimizing reliance on costly conventional and external sources.
\subsection{3-BUS MICROGRID OPERATION WITH OPTIMAL POWER FLOW}
In this case study, the economic dispatch optimization is applied to a 3-bus MG with additional OPF constraints. The OPF analysis explores the spatial and temporal distribution of active and reactive power flows within the microgrid. This analysis incorporates voltage, phase angle, and line flow constraints to ensure stable and reliable operation of the system. In addition, this study evaluates the performance of the system over daily and weekly time horizons.
\subsubsection{Case I: Daily optimal power flow}
\begin{figure*}[t!]
    \centering
    \includegraphics[width = 1\textwidth]{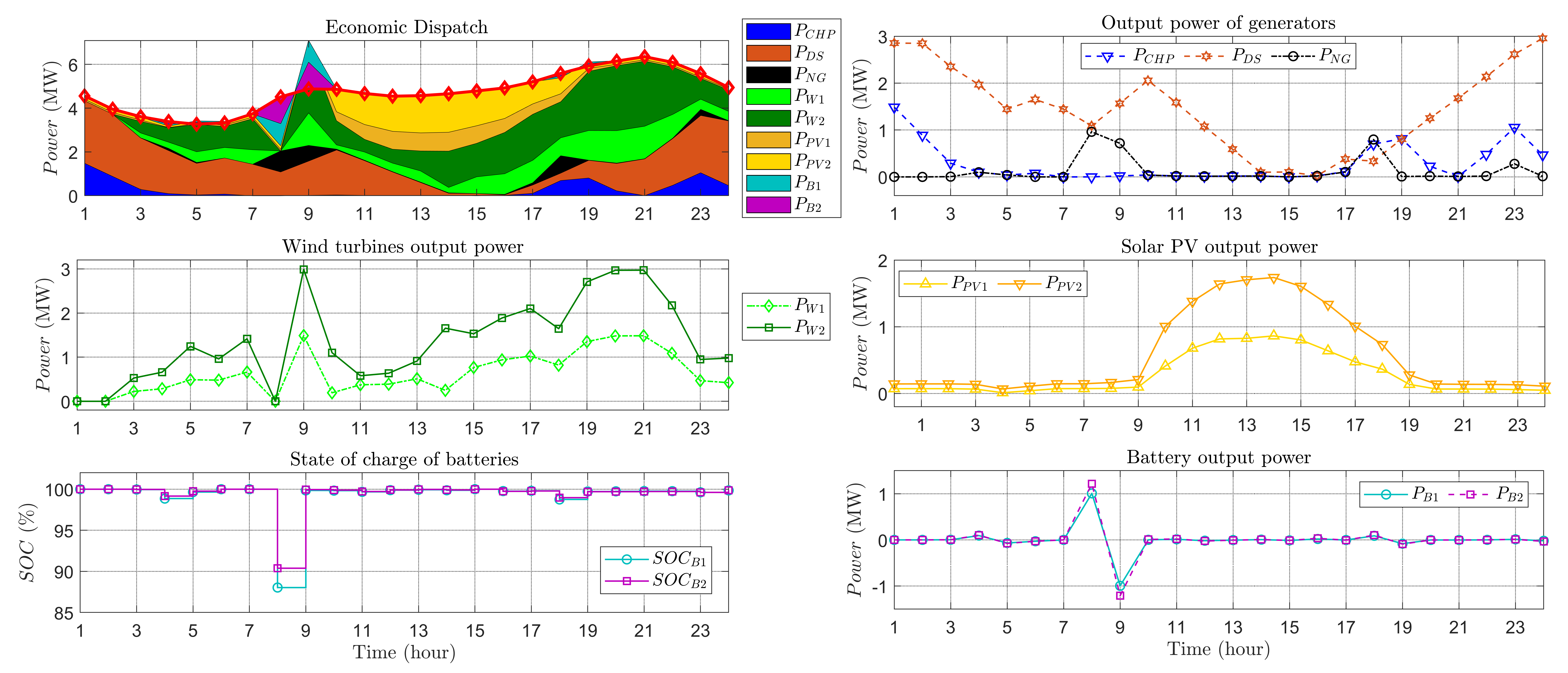}
    \vspace{-0.5cm}
    \caption{Hourly economic dispatch of the 3-bus MG}
    \label{fig:hourlyed}
\end{figure*}

\begin{figure}[t!]
    \centering
    \includegraphics[width = 0.9\columnwidth]{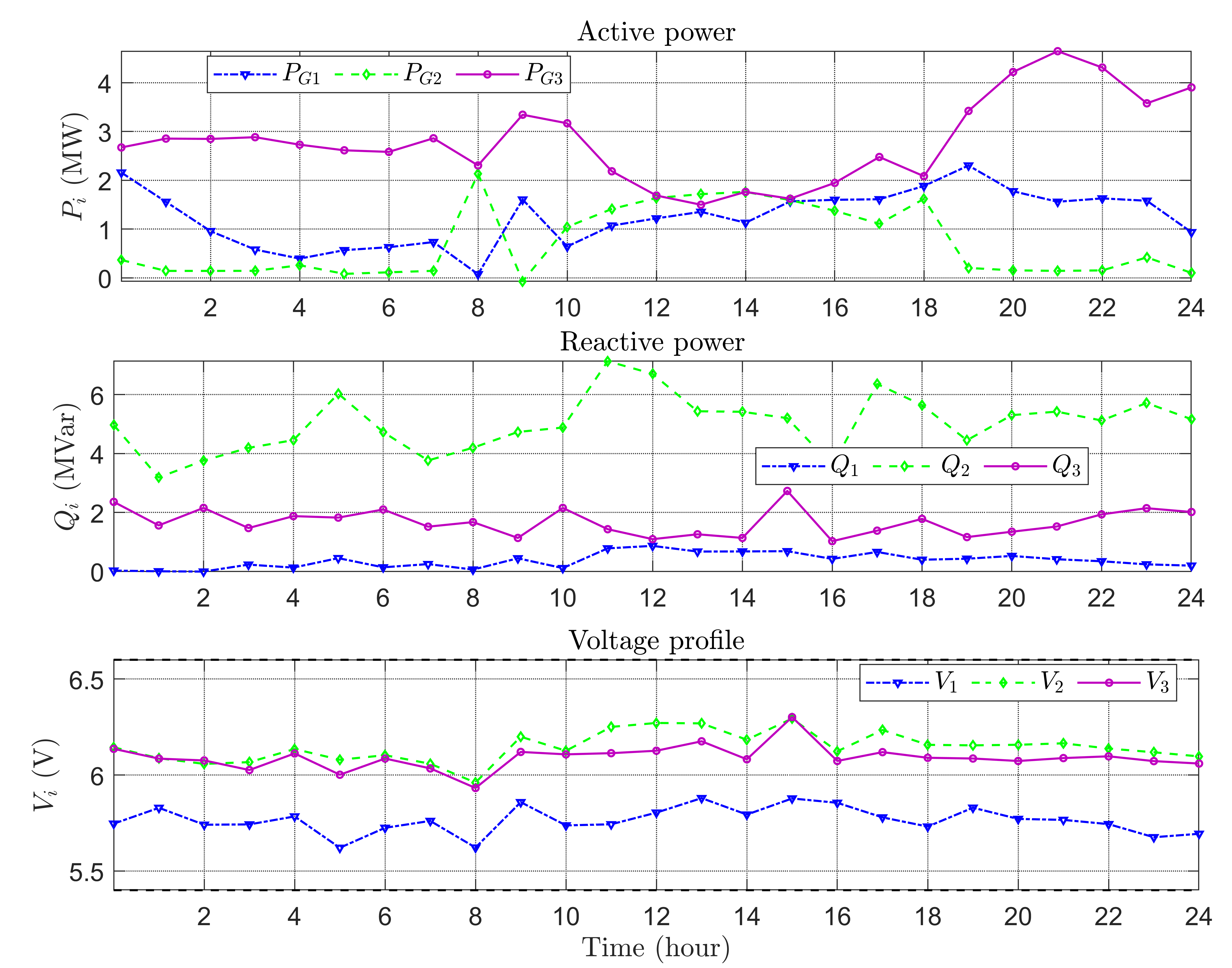}
    \vspace{-0.5cm}
    \caption{Hourly output power and voltage trajectories of the 3-bus MG}
    \label{fig:hourlypqv}
\end{figure}

he results of the daily economic dispatch are illustrated in Fig.\ref{fig:hourlyed}, while the corresponding optimal power flow (OPF) of the 3-bus system is shown in Fig.\ref{fig:hourlypqv}. In this system, the bus active power ($P_i$), bus reactive power ($Q_i$), and bus voltage amplitude ($V_i$) are evaluated to ensure operational efficiency and stability.

As depicted in the top left subplot of Fig.\ref{fig:hourlyed}, the load demand is consistently met throughout the day. The power distribution among the generation units is meticulously adjusted according to their respective cost functions, as detailed in Table\ref{tab:param}. The optimization algorithm ensures that generation from each unit is economically dispatched, balancing cost-effectiveness with system requirements.

The bottom subplot of Fig.~\ref{fig:hourlyed} highlights the limited use of the Battery Energy Storage System (BESS). The BESS is primarily deployed during periods of renewable energy shortfalls, such as when wind turbine output decreases. For instance, at hour 7, the BESS compensates for reduced wind power, maintaining system balance. However, due to its relatively high operational cost compared to other units, the BESS is only employed when absolutely necessary, reflecting its role as a backup resource rather than a primary contributor.
The hourly OPF results are depicted in Fig.\ref{fig:hourlypqv}, where the spatial and temporal distribution of active power ($P_i$), reactive power ($Q_i$), and voltage amplitude ($V_i$) are analyzed. Since the loads dataset does not specify a reactive component, the reactive power loads are assumed to be 20\% of the active power loads. Additionally, the inductive transmission lines, with an impedance of $Z=j1\Omega$, introduce a reactive power component into the system. The dispatch of reactive power from the generators is shown in the middle subplot of Fig.\ref{fig:hourlypqv}.

The top subplot of Fig.\ref{fig:hourlypqv} reveals that Bus 1 contributes the largest share of power among the three buses. This is primarily because the optimization formulation prioritizes cost minimization, favoring the cheapest available power sources to meet both active and reactive power demands. As detailed in Table\ref{tab:param}, Bus 1, which comprises a combination of Combined Heat and Power (CHP) units, wind turbines, and solar PV, can dispatch power at a lower cost compared to the other buses. Consequently, the optimization solver allocates the majority of the generation to Bus 1, demonstrating the effectiveness of the cost-based dispatch strategy.
\subsubsection{Case II: Weekly optimal power flow}
\begin{figure*}[t!]
    \centering
    \includegraphics[width = 1\textwidth]{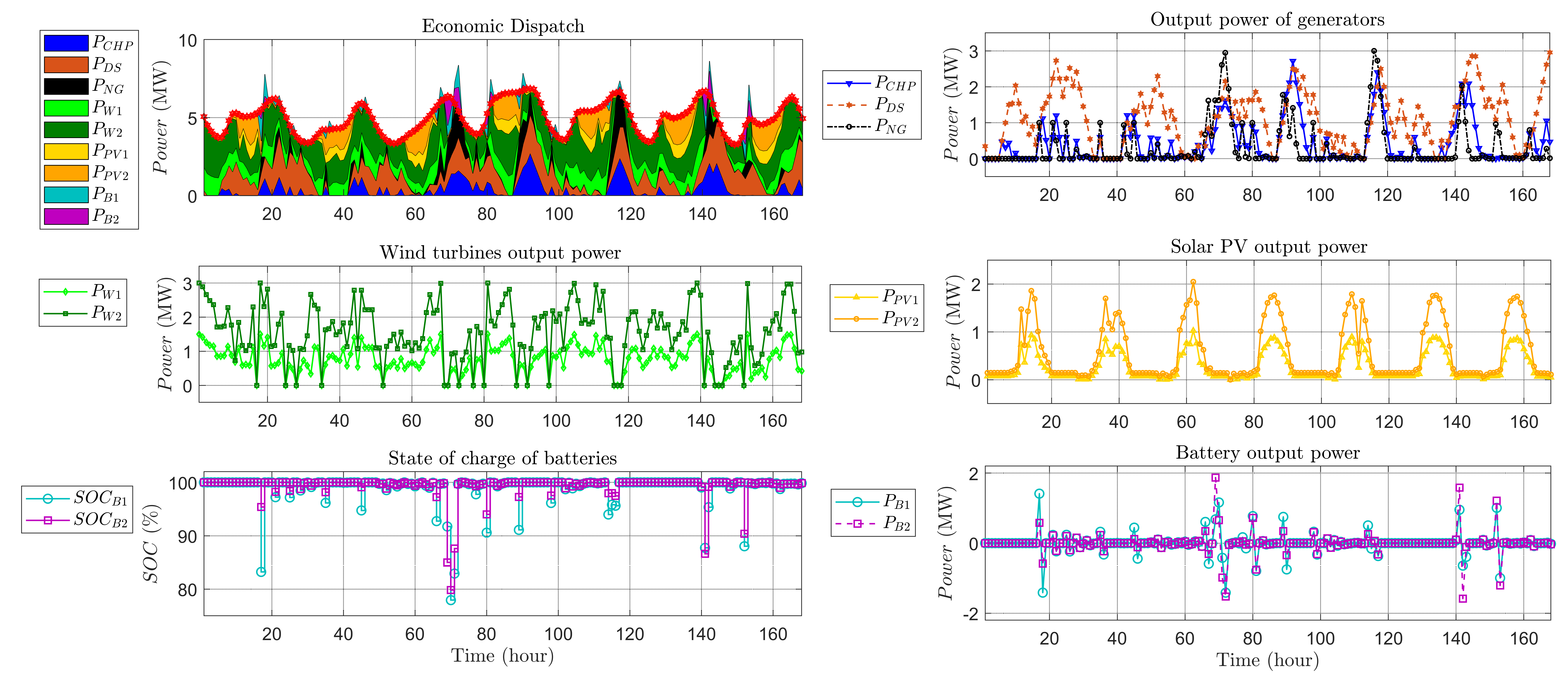}
    \vspace{-0.5cm}
    \caption{Results of the weekly economic dispatch of the 3-bus MG}
    \label{fig:weeklyed}
\end{figure*}

\begin{figure}[ht!]
    \centering
    \includegraphics[width = 1\columnwidth]{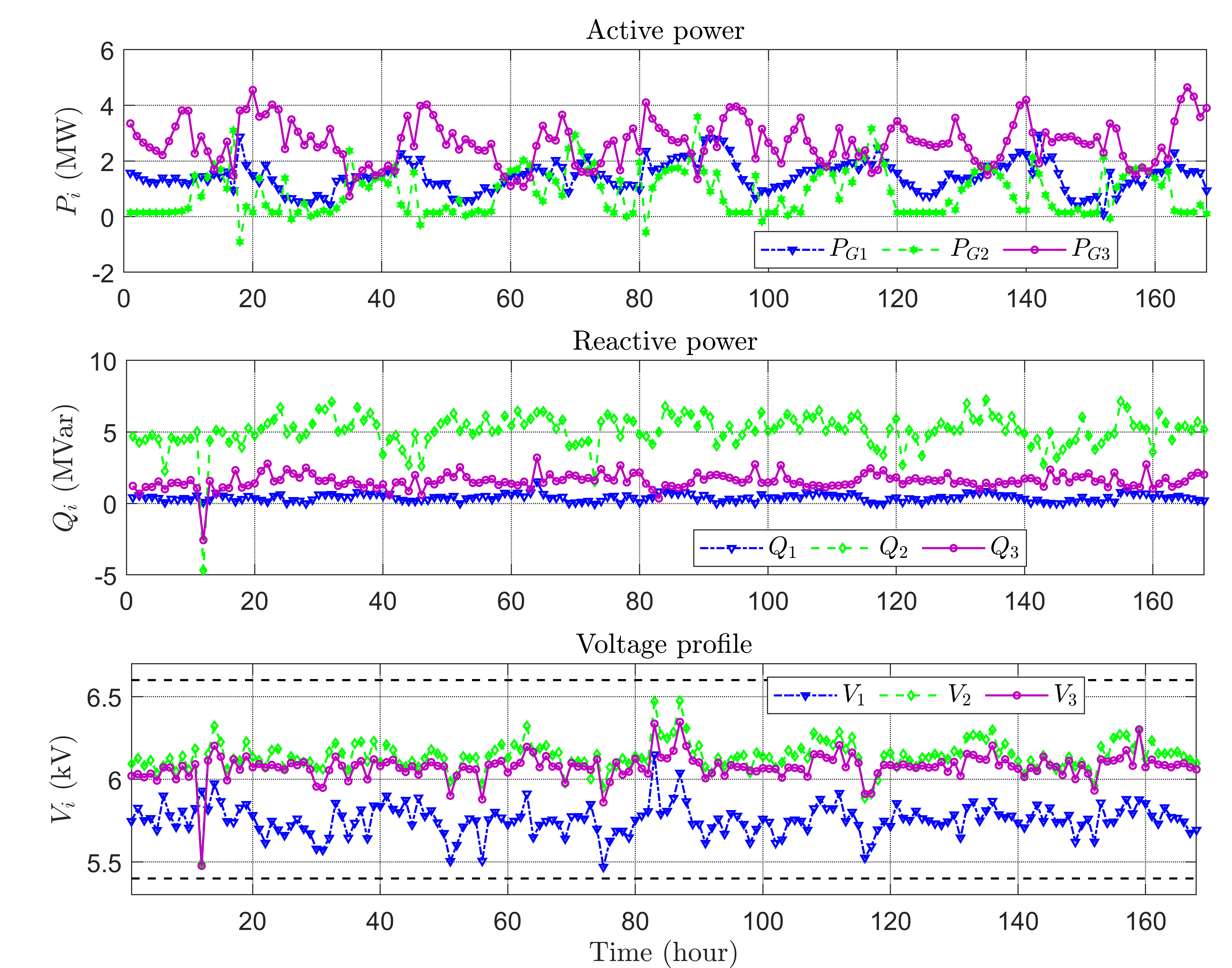}
    \vspace{-0.5cm}
    \caption{Weekly output power and voltage trajectories of the 3-bus MG}
    \label{fig:weeklypqv}
\end{figure}

The weekly economic dispatch extends the simulation to a seven-day period, with the same setup and a total of 168 time steps. The results are presented in Fig.\ref{fig:weeklyed} and Fig.\ref{fig:weeklypqv}, showcasing the system's ability to maintain consistent performance over an extended horizon. Similar to the daily case, the load demand is continuously satisfied throughout the week, as observed in the top left subplot of Fig.~\ref{fig:weeklyed}. Different types of generators contribute varying amounts of power to meet the total demand, reflecting the optimization's dynamic allocation strategy.

Renewable energy sources, such as wind and solar PV, are maximally utilized throughout the week, given their cost-effectiveness compared to conventional generation. This extensive use of renewables is evident in the middle subplot of Fig.\ref{fig:weeklyed} and aligns with the maximum renewables power output shown in (Fig.\ref{fig:PV}-Fig.~\ref{fig:WT}). By prioritizing these cheaper sources, the system minimizes operational costs while leveraging renewable energy's environmental benefits.

In contrast, the Battery Energy Storage System (BESS) plays a critical role in balancing the system by repeatedly charging and discharging throughout the week, as depicted in the bottom subplots of Fig.~\ref{fig:weeklyed}. Notably, the BESS often reaches full charge within an hour, indicating that its current operational constraints may not fully reflect real-world behavior. To address this, it is recommended to refine the optimization formulation by adding constraints on charging and discharging rates. Such constraints would better mimic the practical limitations of BESS operation and enhance system performance under real-world conditions.
The optimal power flow (OPF) results for the weekly case, shown in Fig.\ref{fig:weeklypqv}, highlight the contributions of each bus to meeting load demands. Buses 1 and 3 contribute comparable amounts of power, while Bus 2 has a more modest contribution. This difference can be attributed to the composition of Bus 2, which includes solar PV, batteries, and a natural gas generator. Although solar PV offers a significant cost advantage, its maximum output is restricted to daylight hours. The BESS, despite its critical role, incurs higher operational costs, and the natural gas generator further increases the economic burden relative to other conventional generators. These factors together explain the observed power dispatch pattern, as illustrated in Fig.\ref{fig:weeklyed}.

The voltage trajectories for the weekly case exhibit similar patterns to the daily results, with dips and spikes occurring periodically due to large load changes and switching operations. However, these variations remain within a $\pm 10 \%$ boundary of the nominal 6kV design voltage, as shown in Fig.~\ref{fig:weeklypqv}. While the voltage deviations are within acceptable limits, they indicate potential stability concerns over longer operational periods. To address this, implementing a secondary control mechanism to restore and stabilize voltages could further enhance system reliability and performance.
\section{CONCLUSION}\label{sec:conclusion}
This study presents a comprehensive analysis of economic dispatch and optimal power flow in microgrid systems, addressing both single-bus and three-bus grid-tied configurations. The proposed methodologies ensure cost-efficient operation while maintaining system reliability. Key findings include the maximized utilization of renewable energy sources, the strategic role of BESS in balancing generation and load, and the economic prioritization of conventional and grid power based on cost efficiency. For the weekly horizon, the system demonstrated consistent load satisfaction, with renewable energy sources contributing significantly to cost reduction. However, the BESS's operational behavior highlights the need for enhanced constraints on charging and discharging rates to better align with real-world applications.

The OPF analysis revealed that Bus 1 consistently provided the largest share of power, leveraging its cost-effective combination of renewable and conventional resources. Voltage variations were well-contained within acceptable limits but observed dips and spikes suggest potential stability challenges during dynamic operations. Implementing secondary voltage control mechanisms could further enhance system stability and ensure compliance with operational standards.

Overall, the findings underscore the viability of the proposed optimization framework in achieving sustainable, reliable, and cost-effective microgrid operations.
\section{SOURCE CODE}\label{sec:conclusion}
To facilitate reproducibility and allow readers to explore the methodologies and case studies presented in this paper, the source code for running the simulations and obtaining the data is publicly available. The repository provides all necessary scripts and resources to replicate the economic dispatch and optimal power flow analyses for the single-bus microgrid and the three-bus grid-tied microgrid configurations.

The code can be accessed via the following link:
\texttt{https://t.ly/zv1ZG}

The repository includes:
\begin{itemize}
    \item Input data for load profiles, renewable energy sources, and generator parameters.
    \item Optimization scripts for economic dispatch and OPF formulations.
    \item Post-processing scripts to generate the figures and results discussed in the paper.
    \item Readers are encouraged to use the code to adapt the case studies for their own research or practical applications. 
    \item For any questions or issues, please refer to the documentation in the repository or contact the authors.
\end{itemize}
\bibliographystyle{IEEEtran}
\bibliography{IEEEabrv,ref}

\end{document}